# Interstitial oxygen order and its competition with superconductivity in La$_2$PrNi$_2$O$_{7+\delta}$

Zehao Dong[1]†, Gang Wang[2,3]†, Ningning Wang[2,3]†, Wen-Han Dong[1]†, Lin Gu[4], Yong Xu[1,5], Jinguang Cheng[2,3]\*, Zhen Chen[2,3]\*, Yayu Wang[1,5,6]\*

[1]State Key Laboratory of Low Dimensional Quantum Physics, Department of Physics, Tsinghua University; Beijing 100084, China

[2]Beijing National Laboratory for Condensed Matter Physics, Institute of Physics, Chinese Academy of Sciences; Beijing 100190, China.

[3]School of Physical Sciences, University of Chinese Academy of Sciences; Beijing 100049, China.

[4]School of Materials Science and Engineering, Tsinghua University; Beijing 100084, China.

[5]New Cornerstone Science Laboratory, Frontier Science Center for Quantum Information; Beijing 100084, China.

[6]Hefei National Laboratory; Hefei 230088, China.

†These authors contributed equally to this work.

\*Corresponding author. Email: jgcheng@iphy.ac.cn; zhen.chen@iphy.ac.cn; yayuwang@tsinghua.edu.cn.

**Abstract**

High-temperature superconductivity in bilayer nickelate $La_3Ni_2O_7$ under pressure has attracted significant interest in condensed matter physics[1-4]. While early samples exhibited limited superconducting volume fractions[5,6], Pr substitution for La enabled bulk superconductivity in polycrystals under pressure[7] and enhanced transition temperatures in thin films at ambient pressure[8-10]. Beyond rare-earth doping, moderate oxygen or ozone annealing improves superconductivity by mitigating oxygen vacancies[11,12], whereas high-pressure oxygen annealing leads to a trivial, non-superconducting metallic state across all pressure regimes[13]. These findings highlight the need to elucidate both the individual and combined effects of Pr doping and oxygen stoichiometry in modulating superconductivity in bilayer nickelates. Here, using multislice electron ptychography and electron energy-loss spectroscopy, we investigate the structural and electronic properties of as-grown $La_2PrNi_2O_7$ and high-pressure-oxygen-annealed $La_2PrNi_2O_{7+\delta}$ polycrystals. We find that Pr dopants preferentially occupy outer La sites, effectively eliminating inner-apical oxygen vacancies and ensuring near-stoichiometry in as-grown $La_2PrNi_2O_7$ that is bulk-superconducting under pressure. In contrast, high-pressure oxygen annealing induces a striped interstitial oxygen order, introducing quasi-1D lattice potentials and excess hole carriers into $p$-$d$ hybridized orbitals, ultimately suppressing superconductivity. This behavior starkly contrasts with cuprate superconductors, where similar interstitial oxygen ordering enhances superconductivity instead[14]. Our findings reveal a competition between striped interstitial oxygen order and superconductivity in bilayer nickelates, offering key insights into their distinct pairing mechanisms and providing a roadmap for designing more robust superconducting phases.

**Main Text**

The nickelate superconductor $La_3Ni_2O_7$ is a layered $3d$ transition-metal oxide with a bilayer Ruddlesden-Popper (RP) structure and a relatively high critical temperature ($T_C$) under pressure[1], closely resembling cuprate superconductors[15,16]. Each bilayer block consists of a LaO-NiO$_2$-LaO-NiO$_2$-LaO stacking sequence, while adjacent blocks are separated by a vacant interstitial layer[1]. However, its distinctive $3d^{7.5}$ electronic configuration, involving contributions from both the Ni $d_{z^2}$ and $d_{x^2-y^2}$ orbitals, may give rise to orbital-selective behavior and self-doping effects not observed in cuprates[17-20]. As a result, $La_3Ni_2O_7$ has emerged as a model system for investigating unconventional superconductivity and guiding the design of new materials with enhanced $T_C$.

At ambient pressure, $La_3Ni_2O_7$ crystallizes in an orthorhombic structure with distorted $NiO_6$ octahedra[1], as illustrated in Fig. 1a-b. Superconductivity emerges under high pressure, with an onset temperature near 80 K, accompanied by a structural phase transition[1,7,21,22]. Isovalent Pr doping enables bulk superconductivity in $La_2PrNi_2O_7$ under high pressure, providing an ideal platform for exploring the key parameters governing superconducting (SC) phase. Among these, annealing conditions play a crucial role: annealing in a moderately oxidizing environment enhances the SC phase[8–10,19], whereas annealing under high oxygen pressure results in a trivial metallic phase[13]. The sensitivity of the SC phase to oxygen stoichiometry and spatial distribution mirrors observations in single-layer RP-phase cuprates and nickelates, where non-stoichiometric oxygen atoms occupy various lattice or interstitial sites, modulating hole density and thereby influencing both charge order and superconductivity[23–25]. Understanding the evolution of oxygen defects during annealing is therefore essential for optimizing nickelate superconductors and elucidating their pairing mechanism.

Multislice electron ptychography (MEP), leveraging four-dimensional scanning transmission electron microscopy (4D-STEM) datasets, has emerged as a powerful technique for probing atomic-scale oxygen defects. It offers exceptional sensitivity to light elements and superior robustness across diverse experimental conditions compared to conventional imaging methods[26,27]. Previously, MEP revealed a substantial presence of oxygen vacancies in parent $La_3Ni_2O_7$ single crystals with limited SC volume fractions, underscoring the necessity of post-annealing to enhance the SC phase[12]. Recent advances in ultrafast detectors have significantly reduced the acquisition time of 4D-STEM datasets[28], paving the way for wide-field MEP imaging to facilitate statistical defect analysis and phase-separation studies. Here, we integrate advanced MEP imaging with electron energy-loss spectroscopy (EELS) to further investigate the evolution of oxygen defects and their influence on superconductivity in as-grown $La_2PrNi_2O_7$ (denoted as AG) and high-pressure-oxygen-annealed $La_2PrNi_2O_{7+\delta}$ (denoted as HPO).

**X-ray diffraction and transport properties**

The polycrystalline AG and HPO samples were synthesized using a sol-gel method[7] with protocols detailed in the Methods section, where the HPO samples were post-annealed under an oxygen pressure of 200 bar. The powder X-ray diffraction (XRD) patterns in Fig. 1c reveal that both samples retain an orthorhombic structure at ambient conditions, as evidenced by the splitting of the (020) and (200) diffraction peaks. Notably, the diffraction peaks of the HPO sample shift to smaller angles, corresponding to a ~0.6% expansion in the *a*- and *b*-axis lattice constants. No secondary phases or structural deviations are observed, confirming the preservation of the bilayer RP structure during the annealing process.

Temperature-dependent resistivity measurements reveal pronounced differences between the AG and HPO samples. At ambient pressure (Fig. 1d), AG displays semiconducting behavior, whereas HPO exhibits metallic behavior at high temperatures with a resistivity upturn below 30 K. Figure 1e shows that under a hydrostatic pressure of 14 GPa, AG undergoes a SC transition with onset temperature of 78 K and achieves zero resistance at 40 K (dashed line, adapted from Ref.[7]). In contrast, HPO maintains its metallic behavior across the entire pressure range investigated, with the low-temperature resistivity upturn gradually replaced by a slight resistivity drop. Notably, no SC state is observed in HPO even under pressures up to 14 GPa, consistent with the previous report of suppressed superconductivity in HPO-annealed bilayer nickelate without Pr substitution[13].

**Significant reduction of oxygen vacancies in as-grown La$_2$PrNi$_2$O$_7$**

We then use MEP and EELS to investigate the AG polycrystals, which exhibit a SC volume fraction of ~97%[7]. Figure 2a presents a large-area (35×10 nm$^2$) MEP image along the [110] axis, clearly resolving the characteristic bilayer RP structure, including the oxygen sublattices. The inner apical sites are almost fully occupied across the entire field of view, marking a significant improvement over previous reports on La$_3$Ni$_2$O$_7$ single crystals without Pr substitution[12]. The extracted oxygen map in Fig. 2b further corroborates this finding (see Methods for details): although sparse inner apical vacancies persist (indicated by reddish squares), the majority of oxygen sites remain free of deficiencies. Statistical analysis of the oxygen occupancy (Fig. 2c) estimates the inner apical oxygen vacancy content to be $\delta = 0.07\pm0.15$, reinforcing the observation of a substantial reduction in vacancy concentration, as is also reflected in neutron diffraction refinements[7].

In single-layer RP-phase oxides, previous studies have established that isovalent substitution with smaller cations, such as Pr substitution of La, increases the interstitial volume and accommodates higher oxygen concentrations[29–31]. Figure 2d displays elemental EELS mapping of a representative region, revealing that Pr dopants preferentially occupy La sites adjacent to interstitial layers—a trend consistently observed across multiple regions and in lightly Pr-doped samples (Extended Data Fig. 1). This targeted dopant placement is favorable for maximizing the interstitial volume of the bilayer structure, enabling an even higher oxygen content under identical growth conditions compared to that achievable with a uniform distribution of Pr substitutions. The dramatic reduction of inner apical oxygen vacancies achieved through Pr substitution—directly visualized here—resolves a key puzzle underlying the enhanced SC performance of La$_2$PrNi$_2$O$_7$ relative to its undoped counterpart.

**Interstitial oxygen order in HPO-annealed La$_2$PrNi$_2$O$_{7+\delta}$**

To elucidate the suppression of superconductivity after high-pressure oxygen annealing, we compare the MEP reconstructions of AG and HPO along the [100] axis shown in Fig. 3b and Fig. 3e. Both samples display octahedral tilts consistent with the orthorhombic *Amam* space group, manifested as NiO$_6$ octahedral buckling with reversed tilt patterns between the upper and lower planes within each RP bilayer. However, a critical distinction arises in the interstitial sites between adjacent RP bilayers: AG samples possess vacant interstitial sites (dubbed the pristine 327 phase), while those in HPO samples are periodically occupied by excess oxygen atoms (black arrows in Fig. 3e). These interstitial oxygens repel neighboring apical oxygen atoms, thereby reversing the interlayer tilt alignment along the $c$-axis. Fourier-transform images corroborate this structural reorganization: satellite peaks linked to octahedral tilts transition from a misaligned configuration in AG (Fig. 3c) to a $c$-axis-aligned arrangement in HPO (Fig. 3f). This reversed octahedral tilt pattern signifies a stage-1 interstitial oxygen order, where stage-$n$ denotes ordered interstitial oxygen layers separated by $n$ RP bilayers[32,33,14]. This primary phase retains the in-plane lattice translational symmetry with a periodicity of $b$ (period-$b$). Figures 3g-i display a secondary phase in HPO with in-plane $2b$ periodicity (period-$2b$) and staggered $c$-axis stacking, corresponding to a lower interstitial oxygen content.

Leveraging MEP's high spatial resolution and sensitivity to oxygen atoms, we establish atomic models for the two distinct ordered phases (Fig. 3d and g). Further analysis of the MEP image along the perpendicular [010] axis reveals that the interstitial oxygen atoms lack long-range order along the $a$-axis. Instead, they exhibit spatially varying horizontal stretching, as shown in Extended Data Fig. 2, while octahedral tilts manifest as the slight vertical elongation of planar oxygen columns. This demonstrates that the stage-1 phase retains orthorhombic symmetry (*Amam*-type distortion), consistent with the XRD data (Fig. 1c). The reversed tilts accommodate interstitial oxygen atoms located immediately above the planar oxygen sites, corresponding to fractional coordinates (0.5/0, 0.5, 0.25/0.75) within the unit cell as presented in Fig. 1a. We also statistically determine the Ni-O-Ni bond angles for each planar oxygen (Extended Data Fig. 3). In both period-$b$ and period-$2b$ phases, planar oxygen atoms distorted towards the interstitial sites (O3 in the atomic model of Ref.[7]) exhibit larger bond angles (174.8 ± 1.6°) compared to those distorted away (O4, 167.0 ± 1.9°). In contrast, the pristine 327 phase displays uniformly smaller bond angles (165.4 ± 2.6° for both O3 and O4). Therefore, interstitial oxygen atoms push bond angles closer to 180°, a configuration often associated with enhanced superconductivity[34]. Thus, additional factors must be considered to account for the disappearance of superconductivity in HPO at pressures up to 14 GPa.

**Phase separation and hole doping**

A large-area MEP image (Fig. 4a) reveals that the stage-1 order, characterized by two distinct in-plane periodicities, coexists with the pristine 327 phase in HPO. The region to the right of the white dashed line is predominantly composed of the pristine 327 phase, while the left side is dominated by the stage-1 ordered phase. The latter can be further subdivided into domains exhibiting period-$b$ and period-$2b$ modulations, as demarcated by yellow dashed lines. The period-$2b$ domain preserves an in-plane periodicity of $2b$; however, weak residual interstitial oxygen is detectable within the presumed vacant sites, suggesting incomplete segregation between these coexisting domains. To investigate the structural distortions associated with phase separation, in Fig. 4b we map the Ni-O-Ni bond angles at planar oxygen sites. Here, reddish squares (negative angles) correspond to downward-distorted planar oxygens, whereas bluish squares (positive angles) reflect upward distortions. Notably, the domain boundary between the stage-1 ordered phase and the 327 phase features alternating RP bilayers, where intralayer octahedral distortions reverse polarity after passing through a near-0° distortion (light gray squares). This reversal drives a transition from the aligned interlayer distortion configuration of the stage-1 phase to the misaligned arrangement of the 327 phase. Phase separation along the perpendicular $a$-axis is also supported by depth-dependent slice images reconstructed by MEP (Extended Data Fig. 4).

To assess the carrier doping effect induced by interstitial oxygens, in Fig. 4c we compare the average oxygen $K$-edge EELS spectra of AG and HPO. AG exhibits a prepeak at 529 eV, corresponding to ligand hole states that correlate with oxygen stoichiometry, followed by a dominant peak at 536 eV assigned to high-energy La-$5d$ states, which are independent of oxygen content[12]. The prepeak intensity, normalized to the 536-eV peak, is approximately 0.30—consistent with the stoichiometric 327 phase ($\delta = 0$) reported in Ref.[12], confirming the effective elimination of oxygen vacancies via Pr substitution. In contrast, HPO shows a markedly enhanced prepeak intensity with normalized value exceeding 0.40, accompanied by a redshift in peak energy. According to the established energy band diagram in Ref.[12], these features signify a substantial increase of hole carrier density due to interstitial oxygen doping, which lowers the Fermi energy within the $p$-$d$ hybridized band (Fig. 4d). The large-area prepeak intensity map in Fig. 4e highlights the spatial phase separation: the region in Fig. 4a (white square) shows a domain boundary aligned with the transition from high-intensity prepeak regions (reddish areas, stage-1 ordered phase; intensity $\geq 0.4$) to low-intensity regions (greenish areas, pristine 327 phase; intensity $\leq 0.3$). This mapping also demonstrates that the stage-1 order emerges as the dominant phase in HPO.

**Discussion**

Our MEP and EELS results offer a comprehensive and microscopic perspective on the structural and electronic factors governing superconductivity in bilayer nickelate. We first demonstrate that Pr substitutions preferentially occupy the La sites of the outer rocksalt layers, which promote the elimination of oxygen vacancies and stabilize a near-stoichiometric 327 phase. The marked enhancement of SC volume fraction underscores the pivotal role of inner apical oxygen in bilayer nickelate, which strengthens interlayer superexchange interactions between half-filled, strongly-correlated $d_{z2}$ orbitals[19,11,12,35–37].

The discovery of a stage-1 interstitial oxygen order in the orthorhombic HPO sample reveals a previously unrecognized structural phase in bilayer nickelate. Notably, the suppression of superconductivity under pressures up to 14 GPa establishes this oxygen-ordered state as a competitor to the SC phase. From a structural perspective, density functional theory (DFT) calculations suggest that the quasi-1D interstitial oxygen stripes preserve the orthorhombic octahedral distortions even under high pressure, in sharp contrast to the AG sample (Extended Data Fig. 5). This octahedral tilt configuration should hinder the Ni($d$)-O($p$)-Ni($d$) orbital overlap between Fermi surface electrons, potentially acting as a suppressing factor for superconductivity. Moreover, the orientation of the octahedral distortions (along $b$ axis) coincides with the observed charge and spin density waves[38,39]. As a result, the density waves could be enhanced by the periodic lattice potential, which may persist to high pressure and compete with superconductivity, reminiscent of the "1/8 anomaly" in cuprates[23]. These findings highlight the universality of intertwined orders in strongly correlated quasi-2D superconductors, where subtle structural modulations tip the balance between competing ground states.

Beyond its role in modulating the local atomic potential, interstitial oxygens also directly impact electronic structure through hole doping. Using the linear correlation between MEP-reconstructed phase image and atomic density[12], we quantitatively determine oxygen hyper-stoichiometry through the analysis of phase contrasts between interstitial and planar oxygen columns. This approach yields an oxygen content of 7.23±0.07 for the stage-1 period-$b$ phase (Extended Data Fig. 6), in agreement with values reported for stage-ordered single-layer nickelates[33]. Assuming each interstitial oxygen contributes two holes, we estimate additional doping levels of ~0.23 holes/Ni in the period-$b$ phase and ~0.11 holes/Ni in the period-$2b$ phase. These values intersect with theoretical investigations on bilayer nickelate via the two-orbital $t$-$J$-$J_\perp$ model[17,18,35–37,40–45]. To theoretically elucidate the effect of hole doping, Extended Data Fig. 7 presents the DFT+$U$-calculated band structures. In the high-pressure tetragonal phase of AG-La$_2$PrNi$_2$O$_7$, the bonding $d_{z^2}$ band exhibits relatively flat dispersions along the Γ-X and Γ-Y directions and crosses the Fermi level, consistent with previous reports on La$_3$Ni$_2$O$_7$ and is potentially linked to the observed high-$T_C$ superconductivity[1,17]. In contrast, Bader charge analysis reveals a pronounced doping level of 0.16 holes/Ni in HPO-La$_2$PrNi$_2$O$_{7.23}$ at ambient pressure, leading to an upward energy shift of the flat bonding $d_{z^2}$ band to above the Fermi level. This band becomes further hole-doped upon applying external pressure, which may represent another factor that is unfavorable for superconductivity. This behavior is fundamentally different from cuprates, where ordered interstitial oxygen in La$_2$CuO$_{4+\delta}$ promotes superconductivity by hole doping the half-filled $d_{x^2-y^2}$ orbitals[14]. This dichotomy underscores the distinct electronic structure of nickelates, where orbital selectivity and self-doping effects dominate over the single-band behavior characteristic of cuprates.

In summary, our experiments elucidate the crucial roles of oxygen stoichiometry and spatial ordering in modulating the SC properties in La$_2$PrNi$_2$O$_{7+\delta}$. In particular, the elimination of inner apical oxygen vacancies is beneficial for superconductivity, whereas interstitial oxygen order is strongly detrimental. The interplay between interstitial oxygen order, orbital-dependent doping and high-pressure superconductivity reveals a complex landscape of intertwined orders in bilayer nickelate that is distinctively different from that in cuprates. These results not only advance our understanding of the unique mechanism for nickelate superconductivity, but also provide a roadmap for optimizing materials design to unlock more robust high-$T_C$ phases.

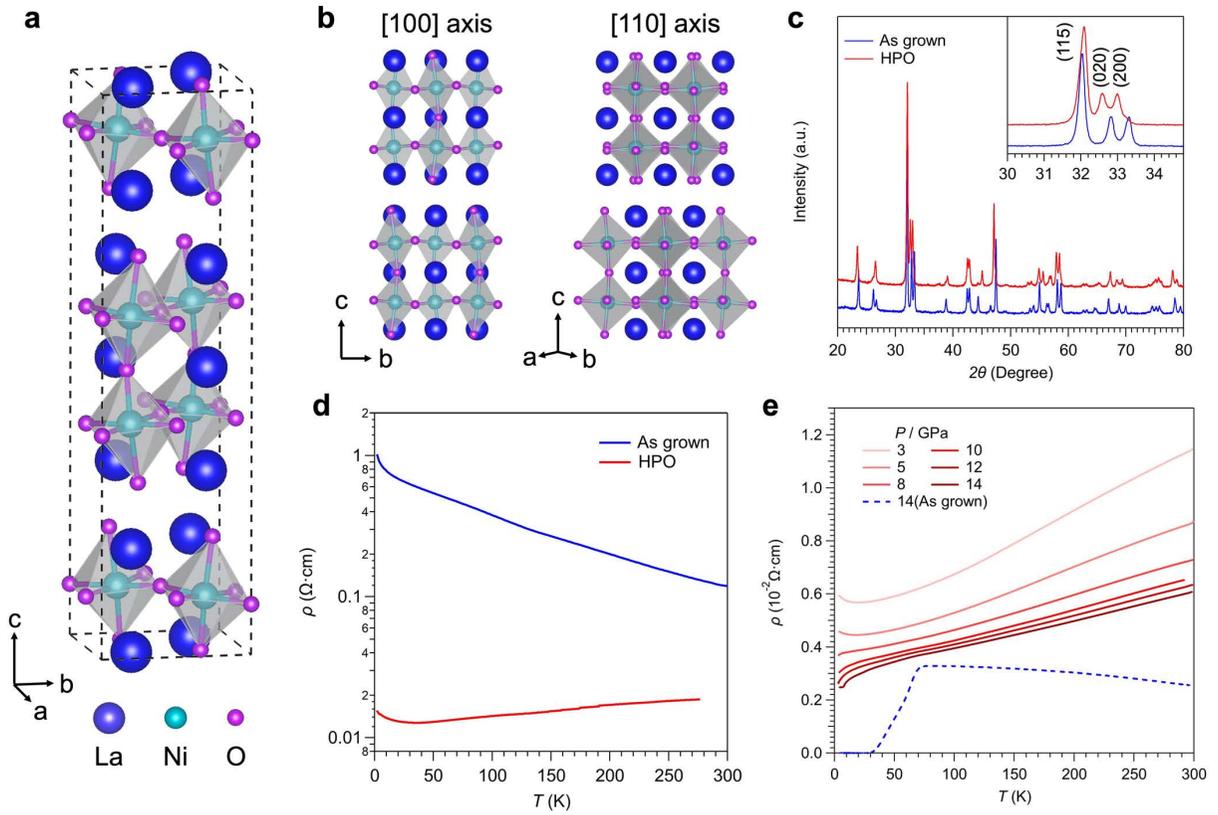

**Figure 1| Characterization of as-grown and HPO-annealed polycrystalline samples. a**, Structural model of the orthorhombic $La_3Ni_2O_7$ unit cell at ambient conditions. **b**, Crystal structure projections along the [100] (left) and [110] (right) crystallographic axes. **c**, Powder XRD patterns for AG and HPO polycrystals. Insets show magnified views of the (020) and (200) diffraction regions. Curves are vertically offset for clarity. **d**, Temperature dependence of resistivity, $\rho(T)$, for AG and HPO samples at ambient pressure. **e**, $\rho(T)$ for the HPO sample under different hydrostatic pressures up to 14 GPa, measured using a cubic anvil cell. The dashed line represents the $\rho(T)$ curve of the AG sample under 14 GPa, reproduced from Ref. [7].

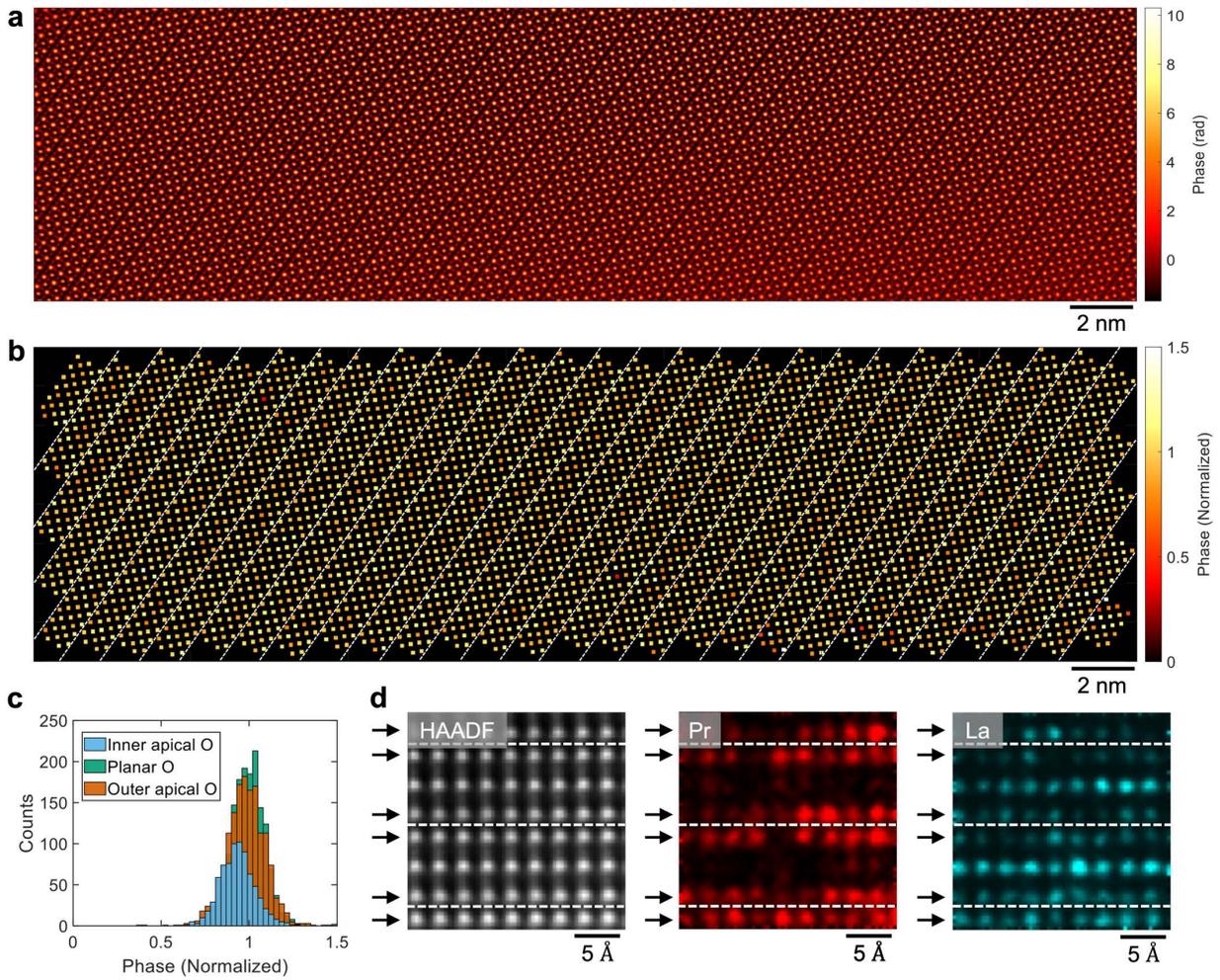

**Figure 2| Atomic structure of the as-grown sample. a**, Projected MEP-reconstructed image acquired across a 35×10 nm$^2$ region of AG. **b**, Normalized oxygen-column map corresponding to panel **a**, color-coded to distinguish oxygen sites. Squares denote individual oxygen columns, with white dashed lines marking boundaries between adjacent RP bilayers. **c**, Histogram of phase distribution for inequivalent oxygen sites in panel **b**. Planar oxygen phases are normalized to unity, revealing inner apical vacancies with a concentration of $\delta=0.07\pm0.15$. **d**, Simultaneously acquired images projected along the [100] axis, including a high-angle annular dark-field (HAADF) image (left panel) and EELS intensity maps of Pr $M_5$-edge and La $M_5$-edge (middle and right panels, respectively). White dashed lines separate adjacent RP bilayers. Black arrows highlight the outer rocksalt LaO layer, where Pr dopants preferentially reside.

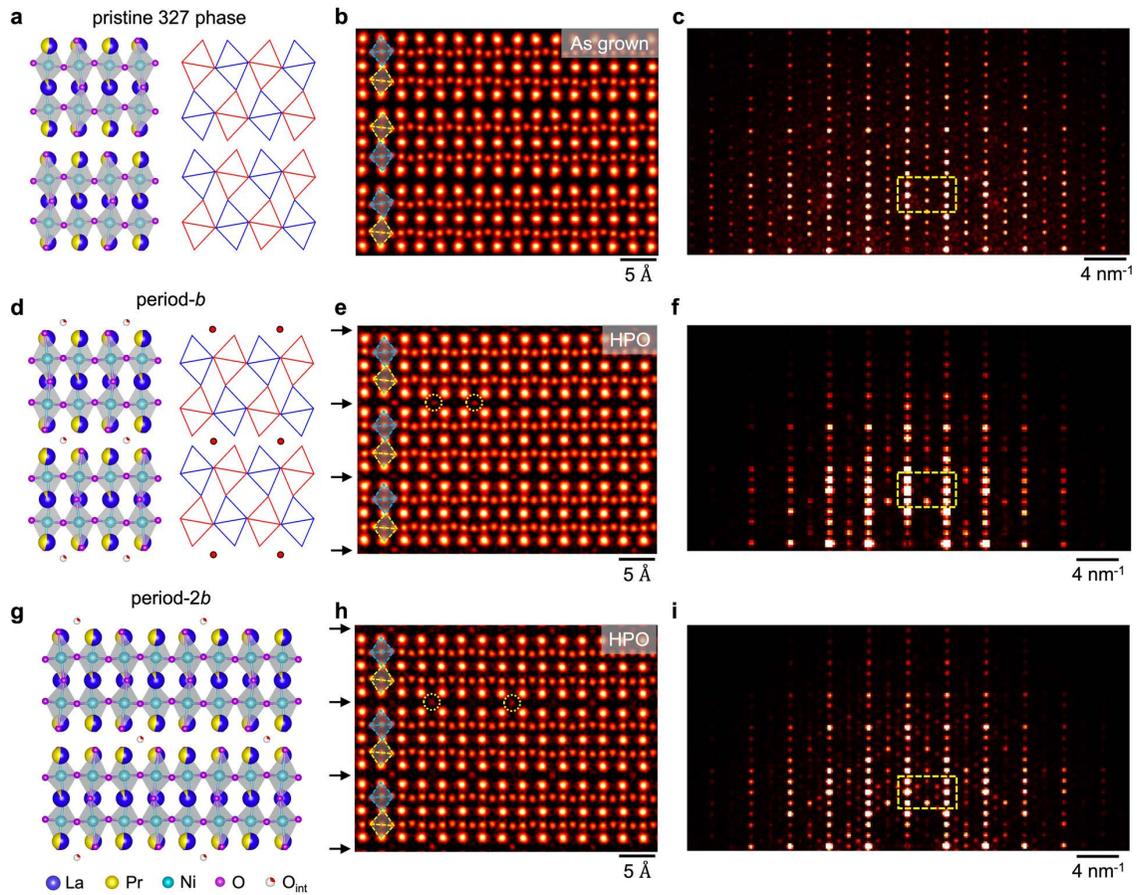

**Figure 3| Interstitial oxygen orders in the HPO-annealed sample. a**, Structural model of AG in the pristine 327 phase (left) and corresponding $NiO_6$ octahedral tilt pattern (right). Clockwise and anticlockwise tilts are depicted as red and blue diamonds, respectively. **b**, Projected MEP-reconstructed image of AG in the pristine 327 phase. Clockwise tilts are represented as yellow diamonds for clarity. **c**, Fourier-transform image of the pristine 327 phase, highlighting misaligned satellite peaks (yellow dashed rectangle). **d**, Structural model of HPO in the stage-1 period-*b* ordered phase. Partially occupied sites are depicted as interstitial oxygen ($O_{int}$) atoms with partial coloring. **e**, Projected MEP-reconstructed image of HPO in the stage-1 period-*b* ordered phase. Interstitial layers and two of the interstitial oxygen columns are marked with black arrows and yellow dashed circles, respectively. **f**, Fourier-transform image of the stage-1 period-*b* ordered phase, showing aligned satellite peaks (yellow dashed rectangle). **g**, Structural model of HPO in the stage-1 period-2*b* ordered phase. **h**, Projected MEP-reconstructed image of HPO in the stage-1 period-2*b* ordered phase. Interstitial layers and two of the interstitial oxygen columns are marked with black arrows and yellow dashed circles, respectively. **i**, Fourier-transform image of the stage-1 period-2*b* ordered phase, revealing additional satellite peaks (yellow dashed rectangle).

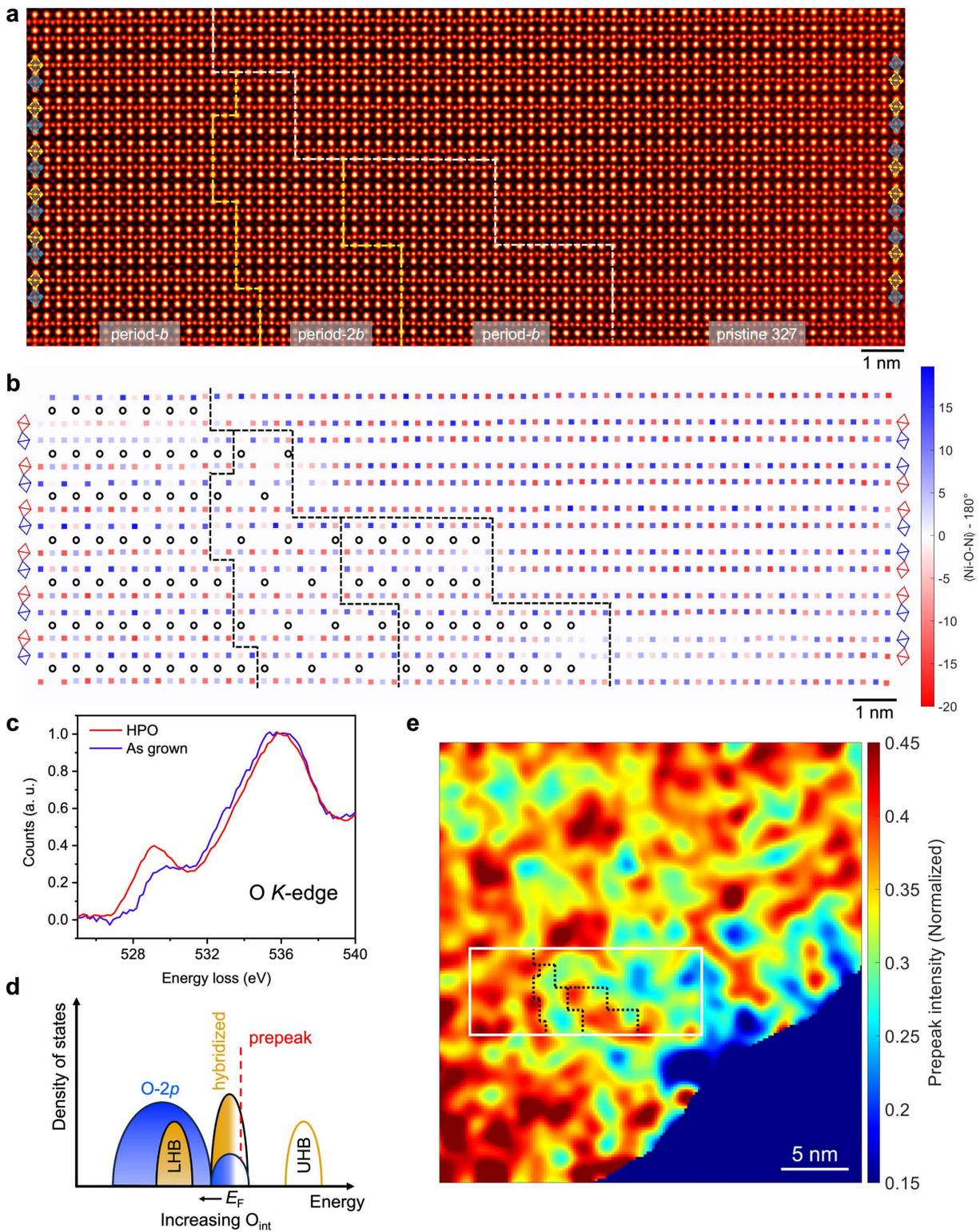

**Figure 4| Phase separation and hole doping from interstitial oxygen order. a**, Large-area MEP-reconstructed image of HPO, displaying the coexisting pristine 327 phase and stage-1

ordered phases. Domain boundaries are demarcated by dashed lines. **b**, Spatial map of planar Ni-O-Ni bond angles derived from panel **a**. Each square represents a planar oxygen, with its color indicating the deviation of bond angle from 180°. Reddish squares (negative angles) correspond to planar oxygens distorted downward, while bluish squares (positive angles) indicate upward distortions. Interstitial oxygen columns are marked as black circles. **c**, Averaged oxygen *K*-edge EELS spectra from HPO and AG samples. **d**, Schematic band structure of bilayer nickelate, illustrating that hole doping via interstitial oxygen lowers the Fermi energy. **e**, Large-area spatial map of prepeak intensity. Region in panel **a** is outlined by white rectangles, while domain boundaries are marked with black dashed lines. The bottom-right vacuum region serves as reference for spatial alignment with panel **a** (0.1-nm precision). Simultaneous HAADF-STEM image and thickness map are provided in Extended Data Fig. 8.

# Methods

## Sample growth and characterization

Polycrystals of La$_2$PrNi$_2$O$_7$ were synthesized using a sol-gel method. Stoichiometric amounts of La$_2$O$_3$, Pr$_6$O$_{11}$ and Ni(NO$_3$)$_2$·6H$_2$O (purity of 99.99%, Alfa Aesar) were dissolved in deionized water, with citric acid and nitric acid added in appropriate amounts. The resulting solution was stirred in a 90 °C water bath for approximately 4 hours to form a vibrant green nitrate gel. This gel was heat-treated overnight at 800 °C to eliminate excess organic matter. The product was then ground, pressed into pellets, and sintered in air at 1100–1150 °C for 24 hours. HPO samples were post-annealed under high oxygen pressure of 200 bar at 500-600 °C for 100 hours.

Powder X-ray diffraction (XRD) data were collected at room temperature via a Huber diffractometer with Cu $K_\alpha$ radiation ($\lambda$=1.5406 Å).

## Transport measurements

Transport properties at ambient pressure were measured using a Quantum Design Physical Properties Measurement System (QD-PPMS) from 2 to 300 K.

A cubic anvil cell (CAC) was used to measure resistivity of AG and HPO polycrystals under pressures up to 14 GPa, with glycerol as the liquid pressure-transmitting medium. The pressure values inside the CAC were estimated using pressure-loading force calibration curves, which were determined by measuring the structural phase transitions of Bi, Sn, Pb, ZnS and GaAs at room temperature.

## Scanning transmission electron microscopy

The transmission electron microscopy (TEM) samples were prepared using a standard powder dispersing method. Polycrystalline powders were ground into an acetone suspension and then deposited onto a Mo holey-carbon TEM grid.

MEP imaging for Fig. 3 and Fig. 4 were performed on JEM-ARM300F2 transmission electron microscope equipped with a GATAN K3 camera and STEMx modular, operating at 300 keV. MEP imaging for Fig. 2 was performed on a JEM-ARM200F transmission electron microscope equipped with a DECTRIS ARINA detector with a high read-out speed, operating at 200 keV. Detailed parameters for MEP experiments are provided in Extended Data Table 1.

MEP reconstruction used a multislice extension of ptychography[46]. A drift-correction algorithm was employed to refine the precise scanning positions from each dataset during MEP reconstruction[47]. Partial spatial coherence was accounted for using a combination of the mixed-state algorithm[48,49] and a Gaussian convolution (with width $\sigma_{PSF}$)[50,51]. The update direction and step size ($\beta_{LSQ}$) for each iteration were determined through the least-squares maximum likelihood (LSQML) method[52,53]. A layer-regularization factor was applied to assist in the convergence of reconstructions[27]. Parameters for reconstructions are provided in Extended Data Table 1.

EELS mapping was performed on JEM-ARM300F2 with a GATAN image filtering (GIF) system and a K3 camera. EELS were acquired under dual-EELS mode, with multiple passes and drift correction between adjacent passes. The pre-edge background was subtracted, and the effect of plural scattering was removed using a standard log-ratio method. Detailed parameters for EELS measurements are provided in Extended Data Table 2. The elemental EELS mapping was extracted using the integrated intensity of Pr $M_5$-edge at 931 eV and La $M_5$-edge at 832 eV, respectively.

**Statistics for atomic phases and positions**

Phases and positions for atoms are determined by fitting the MEP-reconstructed image with two-dimensional Gaussian functions, following a previous work (Ref.[12]).

$$f(x,y) = A\exp\left[-\frac{(x-\mu_x)^2}{2\sigma_x^2} - \frac{(y-\mu_y)^2}{2\sigma_y^2}\right] + B$$

The phase of each atom is proportional to parameter $A$, with its position located at coordinate $(\mu_x, \mu_y)$. To determine the oxygen content, the phase is normalized relative to the average phase of planar oxygen within the same unit cell. Each Ni-O-Ni bond angle is determined by the coordinate of the three atoms.

**DFT calculations**

DFT calculations were performed using Vienna *Ab-initio* Simulation Package (VASP)[54] implementing the projector-augmented wave (PAW) method[55]. We adopted the Perdew-Bruke-Ernzerhof exchange correlational functional revised for solids (PBEsol)[56,57]. For both La$_2$PrNi$_2$O$_7$ (AG) and La$_2$PrNi$_2$O$_{7.23}$ (HPO), we froze the 4$f$ electrons of Pr in the PAW pseudopotentials, since they do not contribute to the Fermi surface[58,59]. High-pressure structural relaxations were performed starting from the ambient-pressure crystal structures determined by XRD refinements (for HPO and AG, listed in the supplementary material). The structures were optimized under pressure until the forces on all atoms were less than 0.01 eV Å$^{-1}$. To handle the fractional atomic occupations of La, Pr and interstitial O, we employed the virtual crystal approximation (VCA)[60] for electronic structure calculations and structural relaxations. In all calculations, we adopted a plane-wave energy cutoff of 550 eV and a Monkhorst-Pack $k$-mesh of 18 × 18 × 5. The DFT + $U$ method was applied for treating 3$d$ electrons of Ni, with the effective Hubbard $U$ = 3.5 eV[19].

## Data and code availability

The code and raw data presented in this study will be available according to the links in the published version.

## Methods references

## Acknowledgments

This work was supported by the Basic Science Center Project of NSFC (No. 52388201), the Innovation Program for Quantum Science and Technology (Grant No. 2021ZD0302502), the NSFC Grant (No. 124B2068, No. U22A6005, No. 52273227, No. 12025408, and No. 12404179), the National Key Research and Development Program of MOST (Grant No. 2023YFA1406400, No. 2022YFA1405100, and No. 2023YFA1406100), and Guangdong Major Project of Basic Research, China (Grant No. 2021B0301030003). Y.W. is partially supported by the New Cornerstone Science Foundation through the New Cornerstone Investigator Program and the XPLORER PRIZE. This work used the facilities of the National Center for Electron Microscopy in Beijing at Tsinghua University, Beijing Laboratory of Electron Microscopy at the Institute of Physics, and the Cubic Anvil Cell Station of the Synergetic Extreme Condition User Facility (SECUF, https://cstr.cn/31123.02.SECUF).

## Author contributions

Z.D., G.W., N.W., and W.-H.D. contributed equally to this work. Y.W., Z.C., and J.C. proposed and supervised the research. Z.D. performed the STEM and EELS experiments and processed data. G.W. and N.W. synthesized polycrystalline samples and performed XRD and transport measurements. W.-H.D. and Y.X. performed DFT calculations. L.G. contributed to electron microscope facilities. Z.D., Z.C. and Y.W. wrote the manuscript. All authors discussed the results and implications throughout the investigation and have given approval to the final version of the manuscript.

## Competing interests

The authors declare no competing interests.

## Additional information

**Correspondence and requests for materials** should be addressed to the corresponding authors.

**Extended data figures and tables**

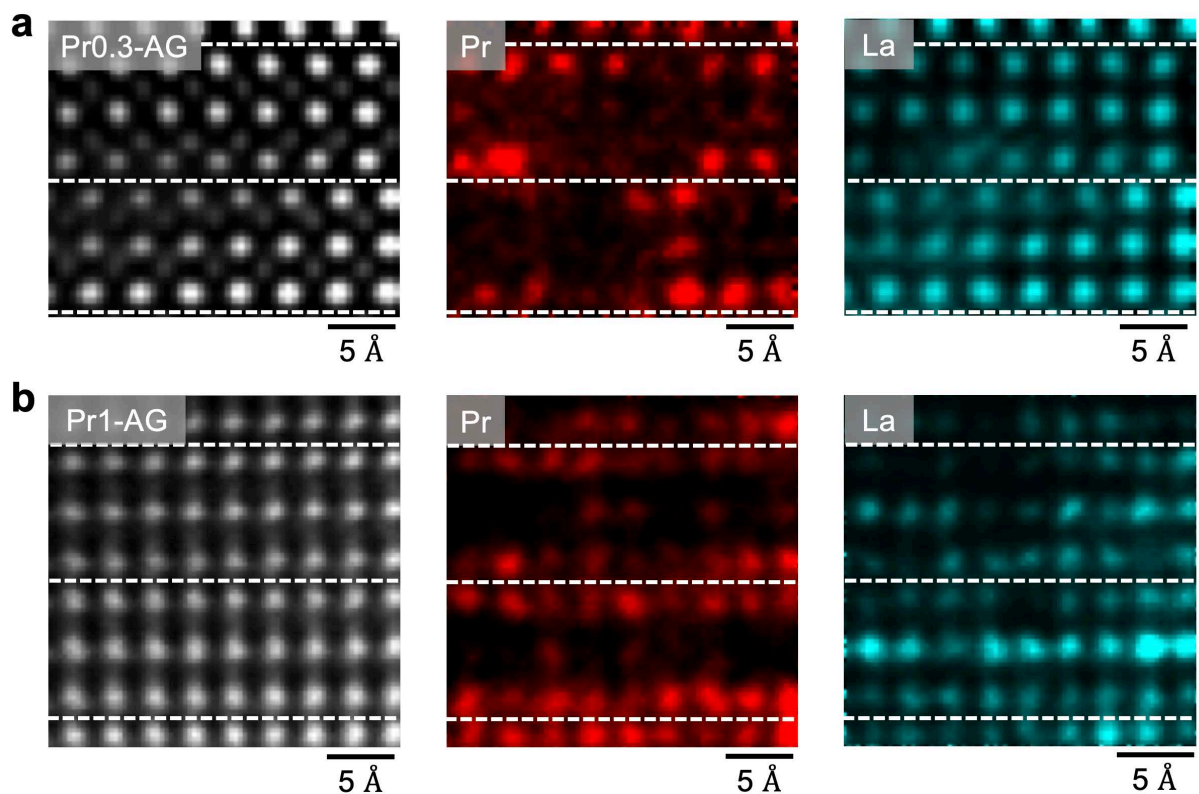

**Extended Data Figure 1| Elemental EELS mapping for Pr dopant distribution. a-b**, Simultaneously acquired HAADF image (left panel), Pr $M_5$-edge intensity map (middle panel), and La $M_5$-edge intensity map (right panel) for as-grown $La_{2.7}Pr_{0.3}Ni_2O_7$ (Pr0.3-AG, panel **a**) and another sample region from as-grown $La_2PrNi_2O_7$ (Pr1-AG, panel **b**). Both regions exhibit the same preference for Pr dopants to substitute La atoms in the outer rocksalt LaO layer.

22**Extended data figures and tables**

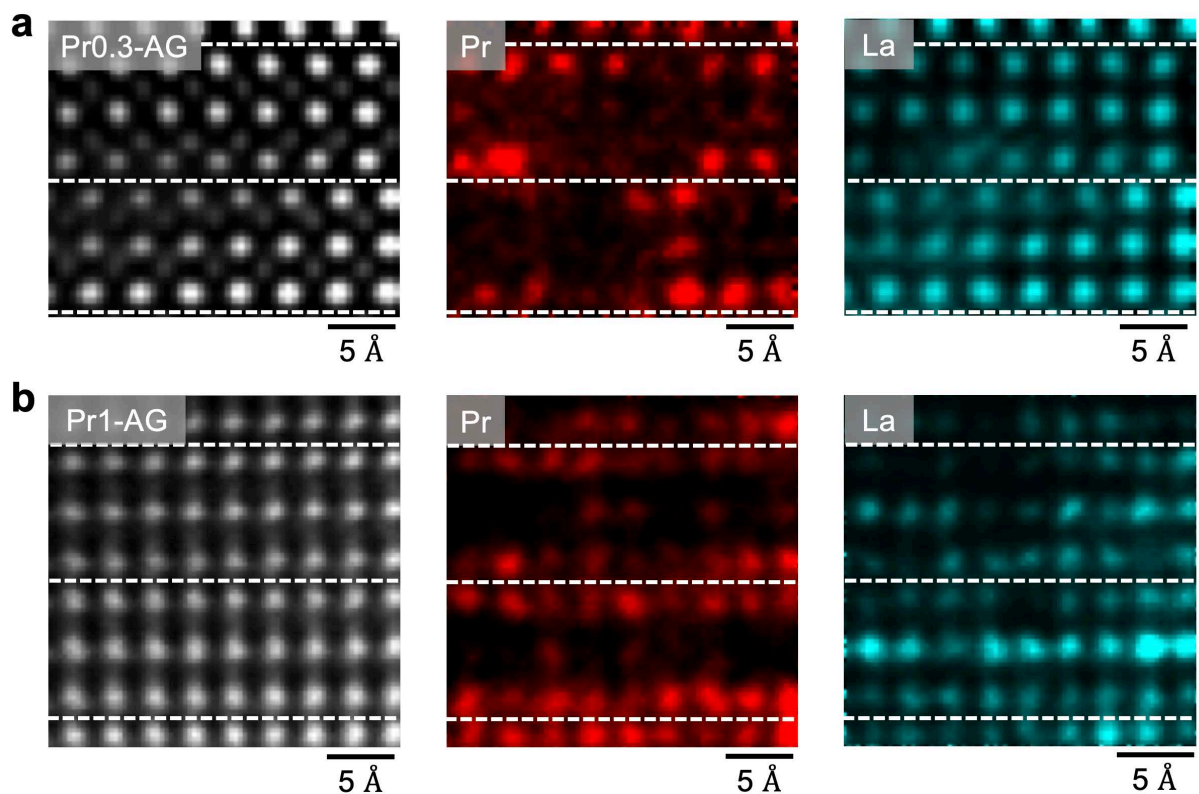

**Extended Data Figure 1| Elemental EELS mapping for Pr dopant distribution. a-b**, Simultaneously acquired HAADF image (left panel), Pr $M_5$-edge intensity map (middle panel), and La $M_5$-edge intensity map (right panel) for as-grown $La_{2.7}Pr_{0.3}Ni_2O_7$ (Pr0.3-AG, panel **a**) and another sample region from as-grown $La_2PrNi_2O_7$ (Pr1-AG, panel **b**). Both regions exhibit the same preference for Pr dopants to substitute La atoms in the outer rocksalt LaO layer.

22**Extended data figures and tables**

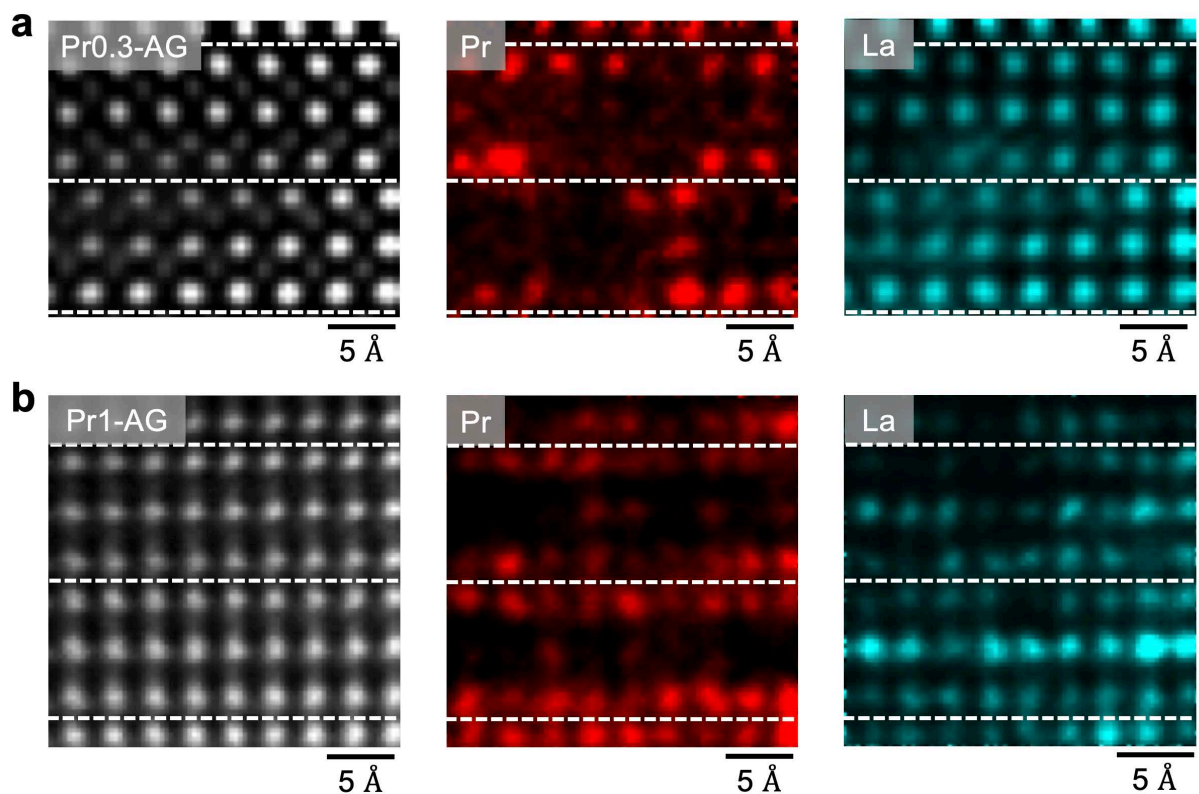

**Extended Data Figure 1| Elemental EELS mapping for Pr dopant distribution. a-b**, Simultaneously acquired HAADF image (left panel), Pr $M_5$-edge intensity map (middle panel), and La $M_5$-edge intensity map (right panel) for as-grown $La_{2.7}Pr_{0.3}Ni_2O_7$ (Pr0.3-AG, panel **a**) and another sample region from as-grown $La_2PrNi_2O_7$ (Pr1-AG, panel **b**). Both regions exhibit the same preference for Pr dopants to substitute La atoms in the outer rocksalt LaO layer.

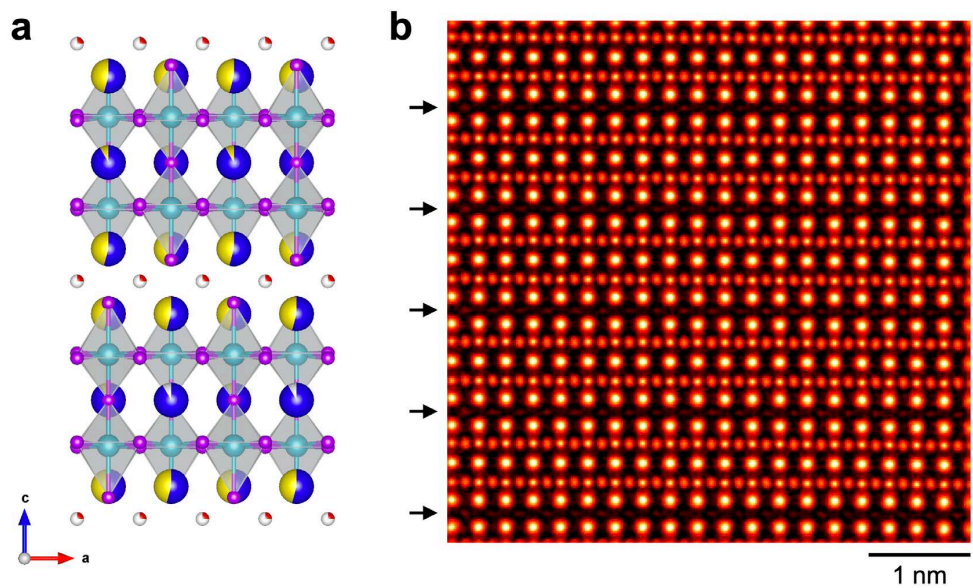

**Extended Data Figure 2| MEP imaging of the HPO-annealed sample along [010] axis. a**, Atomic model for HPO projected along the [010] axis. **b**, MEP reconstructed image of HPO along the [010] axis. Intercalated layers of interstitial oxygens are denoted by black arrows. Note that the octahedral distortions are manifested as the vertical elongation of oxygen columns upon projection.

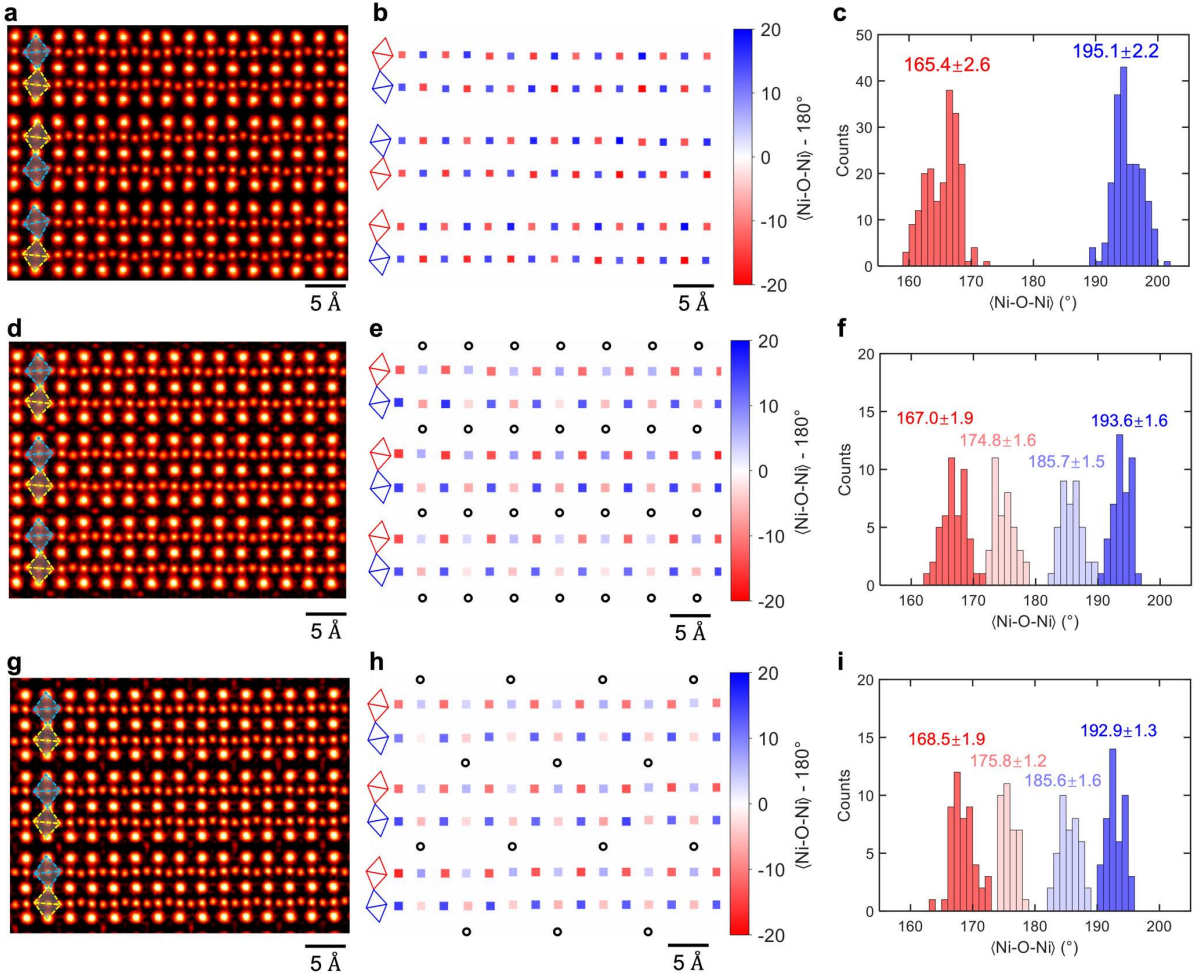

**Extended Data Figure 3| Statistics of Ni-O-Ni bond angles. a**, MEP-reconstructed image corresponding to AG in the pristine 327 phase. **b**, Color-coded map of the planar Ni-O-Ni bond angles of pristine 327 phase. Each square represents a planar oxygen, with its color indicating the deviation of the bond angle from 180°. Reddish squares (negative angles) correspond to planar oxygens distorted downward, while bluish squares (positive angles) indicate upward distortions. **c**, Statistical histogram of planar Ni-O-Ni bond angles for the pristine 327 phase, where the colors of individual components correspond to the respective components in panel **b**. The average bond angles and standard deviation is annotated within the panels. **d**, MEP-reconstructed image corresponding to HPO in the stage-1 period-*b* ordered phase. **e**, Color-coded map of bond angles for the stage-1 period-*b* ordered phase. Interstitial oxygens are represented as black circles. **f**, Statistical histogram of bond angles for the stage-1 period-*b* ordered phase, where the colors of individual components correspond to the respective components in panel **e**. **g**, MEP-reconstructed image corresponding to HPO in the stage-1 period-2*b* ordered phase. **h**, Color-coded map of bond angles for the stage-1 period-2*b* ordered phase. Interstitial oxygens are represented as black circles,

where the colors of individual components correspond to the respective components in panel **g**. **i**, Statistical histogram of bond angles for the stage-1 period-2$b$ ordered phase.

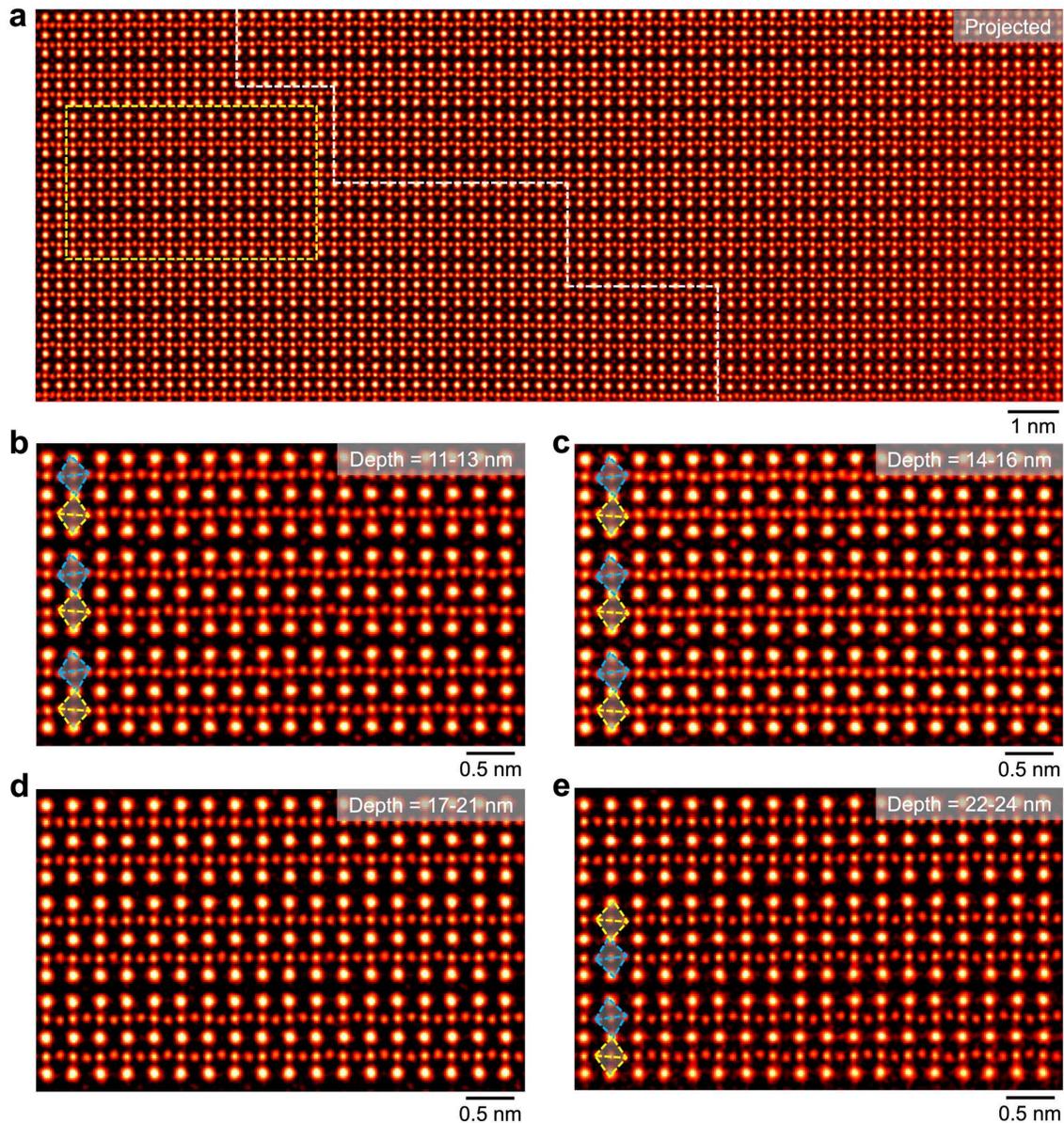

**Extended Data Figure 4| Three-dimensional phase separation in the HPO-annealed sample.
a**, Projected MEP-reconstructed phase image of HPO sample as shown in Fig. 4a. **b-e**, Slice images extracted within the yellow dashed rectangle in panel **A**, from the depth range of 11-13 nm (**b**), 14-16 nm (**c**), 17-21 nm (**d**), and 22-24 nm (**e**), respectively. Panels **b** and **c** display the stage-1 order in the period-*b* phase, while panel **e** exhibits the pristine 327 phase. Panel **d** shows a crossover region between these two phases where octahedral distortions are not well-defined. These results provide clear evidence of a three-dimensional phase separation.

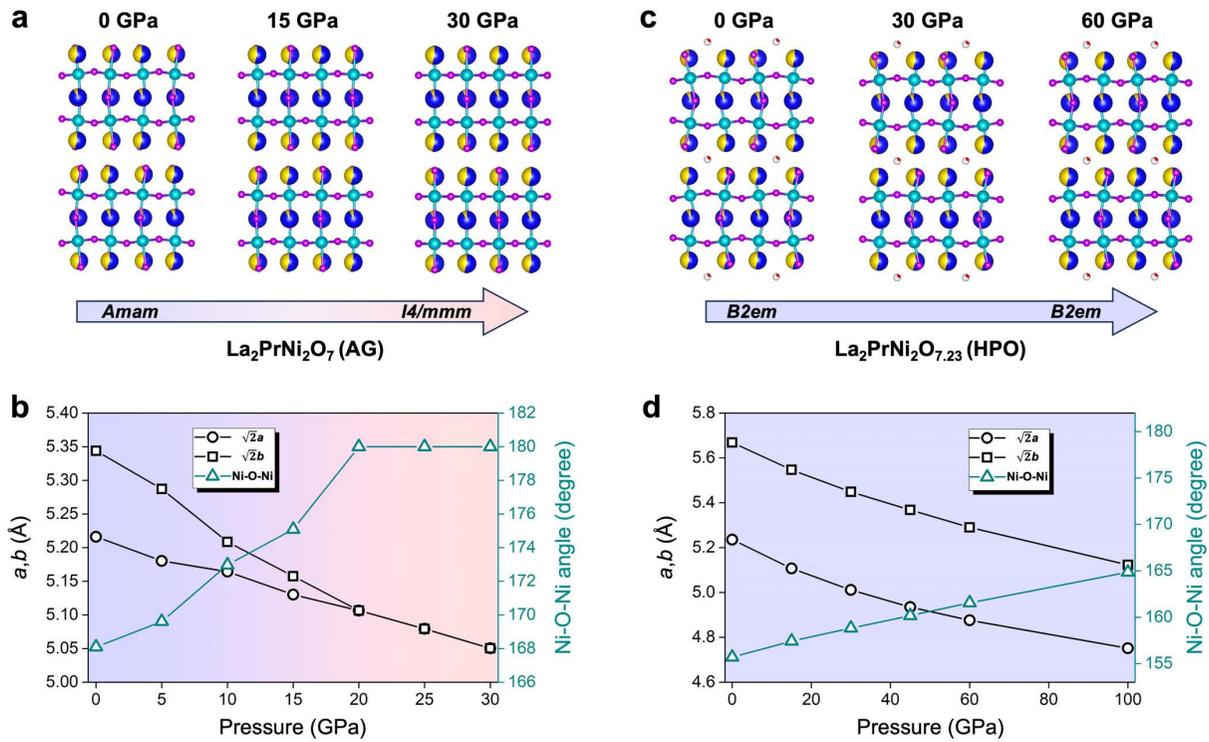

**Extended Data Figure 5| DFT-calculated structural evolutions under pressure. a-b,** Schematics for the relaxed structures (**a**), lattice constants and Ni-O-Ni bond angles (**b**) as a function of the hydrostatic pressure for AG phase $La_2PrNi_2O_7$. The structural phase transition from *Amam* to *I4/mmm* is found around 15 GPa. **c-d,** Schematics for the relaxed structures (**c**), lattice constants and Ni-O-Ni bond angles (**d**) as a function of the hydrostatic pressure for HPO phase $La_2PrNi_2O_{7.23}$. No structural phase transition is observed up to 100 GPa. An effective Hubbard $U = 3.5$ eV was applied when optimizing the crystal structures.

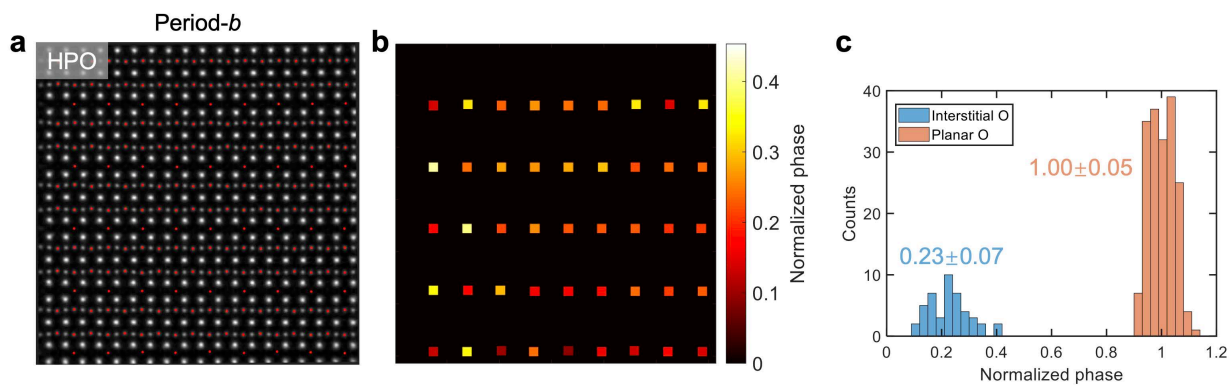

**Extended Data Figure 6| Statistics of interstitial oxygen content of the HPO sample. a**, MEP-reconstructed image from Fig. 3e, with the positions of interstitial and planar oxygens highlighted as red points. **b**, Color-coded map showing the normalized phase values of interstitial oxygens relative to planar oxygens. **c**, Histogram of phase values for interstitial and planar oxygens, showing a normalized phase value of 0.23±0.07 for interstitial oxygen columns relative to planar oxygens. Accordingly, the resulting oxygen hyper-stoichiometry is also estimated as δ = 0.23±0.07.

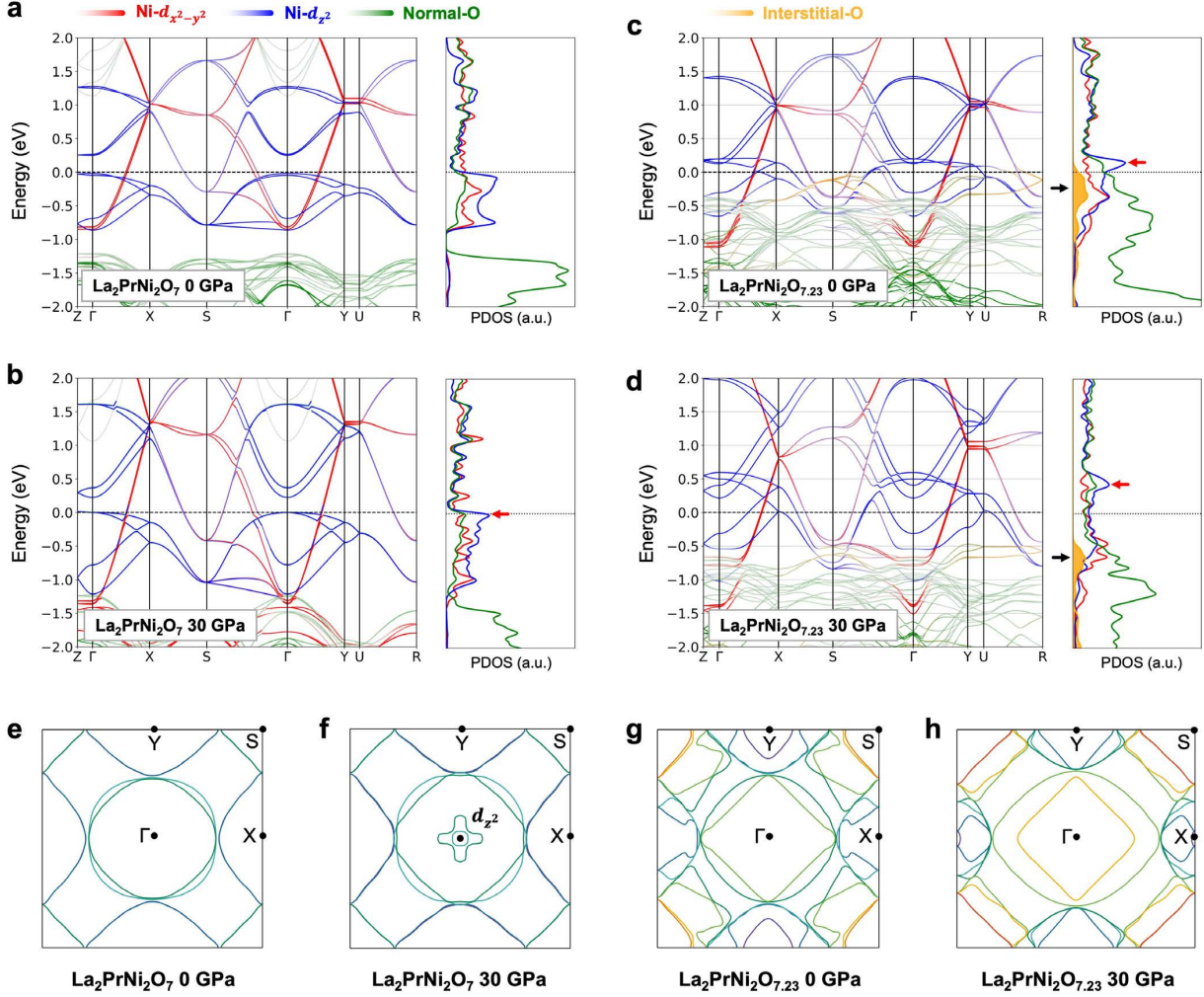

**Extended Data Figure 7| DFT+$U$ calculated electronic structures. a-d,** Projected band structures and projected density of states (PDOS) for AG phase $La_2PrNi_2O_7$ at 0 GPa (**a**) and 30 GPa (**b**), HPO phase $La_2PrNi_2O_{7.23}$ at 0 GPa (**c**) and 30 GPa (**d**), respectively. Experimental crystal structures from Supplementary Materials were used for panels **a** and **c**, respectively. We compared the bands with those from DFT-relaxed structures at 0 GPa, and found no obvious difference. Red arrows indicate the energy of the PDOS peaks at the band edge of bonding $d_{z^2}$ orbitals. Black arrows denote states associated with interstitial oxygens. **e-h**, Two-dimensional Fermi surfaces for panels **a-d,** respectively. Different colors represent distinct band indices.

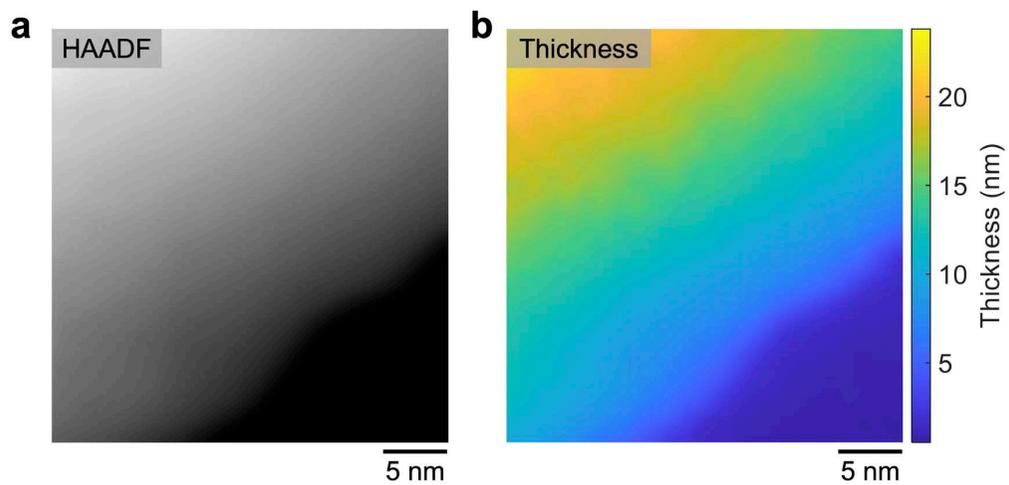

**Extended Data Figure 8| Additional information for the prepeak mapping in Fig. 4e. a**, Simultaneously acquired HAADF image for the mapping area. **b**, Thickness map derived from low energy-loss EELS.

**Extended Data Table 1| Experimental and reconstruction parameters for MEP images**

|  | Fig. 2a | Fig. 3b | Fig. 3e & 3h | Fig. 4a |
|---|---|---|---|---|
| **Microscope** | JEM-ARM200F | JEM-ARM300F2 | | |
| **Camera** | DECTRIS ARINA | GATAN K3 | | |
| **Acc. voltage** | 200 keV | 300 keV | | |
| **Probe semi-angle** | 27.6 mrad | 24.3 mrad | | |
| **Probe current** | 130 pA | 10 pA | | |
| **Dwell time** | 34 μs | 4 ms | | 0.8 ms |
| **Defocus** | Overfocus 20 nm | | | |
| **Diffraction pixels** | 192×192 | 256×256 | | |
| **Diff. binning factor** | 2 | | | |
| **Scan positions** | 1200×600 | 150×150 | 150×150 | 400×150 |
| **Scan step (Å)** | 0.332×0.332 | 0.524×0.524 | | |
| **Diff. sampling (Å$^{-1}$)** | 0.0423×0.0423 | 0.0310×0.0310 | | |
| **Max. diff. angle** | 50.9 mrad | 39.0 mrad | | |
| **No. of iterations** | 500 | 1500 | 2000 | 2000 |
| **Batch size** | 3000 | 200 | 200 | 600 |
| **beta_LSQ** | 1.0 | 1.0 | 1.5 | 1.5 |
| **σ$_{PSF}$ (pixel)** | 1.25 | 1.00 | 1.50 | 1.75 |
| **No. of modes** | 3 | 5 | 2 | 2 |
| **Obj. sampling (Å)** | 0.246 | 0.168 | 0.169 | 0.201 |
| **Layer thickness (Å)** | 6 | 6 | 6 | 10 |
| **Obj. size (pixel)** | 2053×1084 | 791×791 | 779×779 | 1427×668 |
| **No. of slices** | 34 | 35 | 40 | 34 |
| **Layer reg.** | 0.3 | | | |

**Extended Data Table 2| Experimental parameters for EELS measurements**

|  | **Fig. 2d** | **Fig. 4e** |
|---|---|---|
| **Microscope** | JEM-ARM300F2 ||
| **Detector** | GATAN K3 ||
| **Accelerating voltage** | 300 keV ||
| **Convergence semi-angle** | 24.3 mrad ||
| **Probe current** | 40 pA ||
| **Dwell time (ms)×No. of passes** | 0.34×30 | 5×12 |
| **Collection semi-angle** | 77.3 mrad ||
| **Scan pixels** | 60×60 | 164×157 |
| **Scan step (Å)** | 0.4×0.4 | 2×2 |
| **Dispersion (eV/channel)** | 0.18 ||